\documentclass[pra,twocolumn]{revtex4} 
\usepackage{graphicx} 
\usepackage{amsmath, amssymb}
\usepackage{color}

\begin{document}
\title{Supersolid states in a hard-core 
Bose-Hubbard model on a layered triangular lattice}
\author{Ryota Suzuki and Akihisa Koga}

\affiliation{Department of Physics, Tokyo Institute of Technology, 
Tokyo 152-8551, Japan}

\date{\today}

\begin{abstract}
We study ground-state properties in a hard-core Bose-Hubbard model 
on a layered triangular lattice. 
Combining cluster mean-field theory with the density matrix 
renormalization group method, 
we discuss the effect of the interlayer coupling 
on the supersolid states realized in a single layered model.
By examining the distributions for the particle density and 
superfluid order parameter, the rich phase diagram of the system is obtained. 
We find that the supersolid states are widely stabilized 
at a commensurate filling, in contrast to the case of the single layered model.
The nature of the supersolid states is also addressed.
\end{abstract}
\maketitle

\section{Introduction}
Ultracold bosonic systems have attracted much interest
since the successful observation of Bose-Einstein condensation
in $^{87}$Rb atoms~\cite{Rb}. 
One of the interesting topics 
in this field is an optical lattice system~\cite{ol1,ol2,ol3,ol4},
which is formed by loading the ultracold atoms in a periodic
potential. 
This gives us clean correlated bosonic systems on various lattices 
with controllable parameters. 
In fact, remarkable phenomena
have been observed such as the phase transition between
Mott insulating and superfluid states
on cubic~\cite{Greiner} and triangular lattices~\cite{Becker}.
Recently, the possibility of a supersolid state, where
solid and superfluid states coexist, has been 
discussed. 
The existence of the supersolid state has experimentally been suggested 
in the $^4$He system~\cite{Kim}, and has theoretically been discussed in
strongly correlated systems such as 
bosonic~\cite{Matsuda,Mullin,Liu,Ohgoe,Suzuki,Hassan,Suzuki0,Bonnes,Wessel,Yamamoto,Zhang,Boninsegni} 
and fermionic~\cite{Scalettar,Freericks,SSTakemori,SSKoga} systems and
Bose-Fermi mixtures~\cite{Buechler,Titvinidze}.
As for a hard-core bosonic model, it has been clarified that
lattice geometry as well as intersite correlations play a key role
in stabilizing the supersolid state~\cite{Boninsegni,Wessel,Suzuki}.

The hard-core Bose-Hubbard model on a triangular lattice
has been discussed 
in detail~\cite{Wessel,Yamamoto,Boninsegni,Zhang,Bonnes}. 
In the weak coupling region,
the superfluid state is realized.
On the other hand, in the strong coupling region,
the solid state appears at a one-third (two-thirds) filling,
where one of three sites is filled (empty) in a $\sqrt{3}\times\sqrt{3}$
ordering with wave vector ${\bf Q}_2=(4\pi/3,0)$~\cite{Metcalf}.
Between the superfluid and solid states, 
the supersolid state is realized around half-filling, where 
a part of particles crystallize and 
the others form the superfluid state~\cite{Wessel}. 
By contrast, no such supersolid states 
have been studied in the Bose-Hubbard model on layered triangular lattice. 
This should be crucial to observe the supersolid state
in the realistic optical lattice systems~\cite{Martiyanov}.
An important point is that the supersolid state on each layer 
has three-fold degeneracy for particle distributions.
Therefore, it is desired to discuss how 
the interlayer couplings affect the stability of the supersolid state.

\begin{figure}[htb]
\includegraphics[width=8cm]{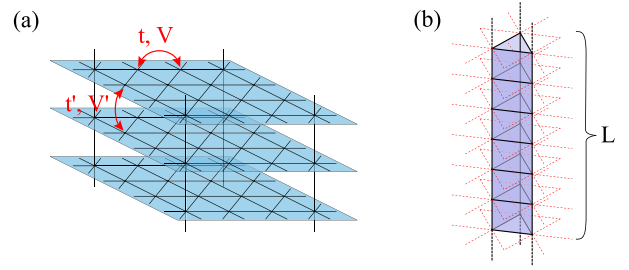}
\caption{(Color online) 
(a) Bose-Hubbard model on layered triangular lattices. 
(b) The effective cluster model with a tube structure, 
which are treated in the framework of cluster mean-field theory.}
\label{fig:layered-model}
\end{figure}

To clarify this point, we systematically study ultracold bosons 
in an optical lattice with the layered triangular structure, 
which is schematically shown in Fig. \ref{fig:layered-model} (a).
We discuss ground-state properties of this model, combining 
cluster mean-field (CMF) theory~\cite{Oguchi}
with the density matrix renormalization group (DMRG) method~\cite{White,DMRG}.
The rich phase diagram is obtained where the solid and supersolid states 
with various particle distributions compete with the superfluid state.
We demonstrate that the supersolid state is realized even at half filling, 
in addition to the well-known supersolid states discussed 
in a single layered model~\cite{Wessel}.
We also discuss the nature of the supersolid states.

The paper is organized as follows. 
In \S\ref{2}, we introduce the Bose-Hubbard model 
for the layered optical lattice system
and explain the possible particle distributions 
for the solid and supersolid states.
We briefly outline our theoretical approach. 
In \S\ref{4}, we study ground-state properties in the layered system
and discuss the stability of the supersolid states. 
A summary is given in the final section.

\section{Model and Method}\label{2}
We consider ground-state properties of interacting bosons 
on the layered triangular lattice. 
Here, we assume sufficiently large onsite interactions. 
In the case, the system should be described by
the following hard-core Bose-Hubbard model as,
\begin{eqnarray}
H&=&  - \sum_{\langle i,j\rangle} t_{ij}(a_i^\dagger a_j + h.c.)
  + \sum_{\langle i,j\rangle} V_{ij}n_i n_j
  \label{hardcore-bose-hubbard}
\end{eqnarray}
where $\langle i, j \rangle$ denotes the summation over nearest neighbor sites 
and $a_i^{\dagger} (a_i)$ 
is the creation (annihilation) operator of a boson at site $i$.
The hopping $t_{ij}=t\; (t')$ and the intersite interaction $V_{ij}=V\; (V')$ 
when the sites $i$ and $j$ are located in the same (distinct) layer.
This model is symmetric when interchanging particles with holes.
The symmetric condition is given by $\mu/V = 3 + r$,
where $\mu$ is the chemical potential and $r=V'/V$.
In the paper, we restrict our discussions to the system with $\mu/V \le 3+r$,
where the particle density $n(=\sum_i\langle n_i\rangle/N)\le 1/2$ 
and $N$ is the total number of sites.

When the interlayer couplings are zero $(t'=V'=0)$,
the model Hamiltonian eq. (\ref{hardcore-bose-hubbard}) is reduced to 
the hard-core Bose-Hubbard model on a single-layer triangular 
lattice~\cite{Wessel}.
In the limit, it is known that the supersolid states are realized, 
which are surrounded by the superfluid state in the weak coupling region and 
solid states with $n = 1/3$ and $2/3$ 
in the strong coupling region~\cite{Wessel}.
Note that at the particle-hole symmetric case, 
two supersolid states with distinct particle density are degenerate.
Therefore, when the system is fixed at half filling, 
the genuine supersolid state is not stable, but
the phase separation between two supersolid states appears 
in the strong coupling region.
When $t=V=0$, the model is reduced to 
the one-dimensional Bose-Hubbard model,
where no supersolid state appears~\cite{Cazalilla,YangYang}.
In the paper, we study the hard-core Bose-Hubbard model 
on the layered triangular lattice 
to discuss how the supersolid states is realized.

First, we define the order parameters for 
the possible ground states.
In the layered system with intersite interactions, 
various type of particle distributions are naively expected.
\begin{figure}[htb]
  \centering
  \includegraphics[width=8cm]{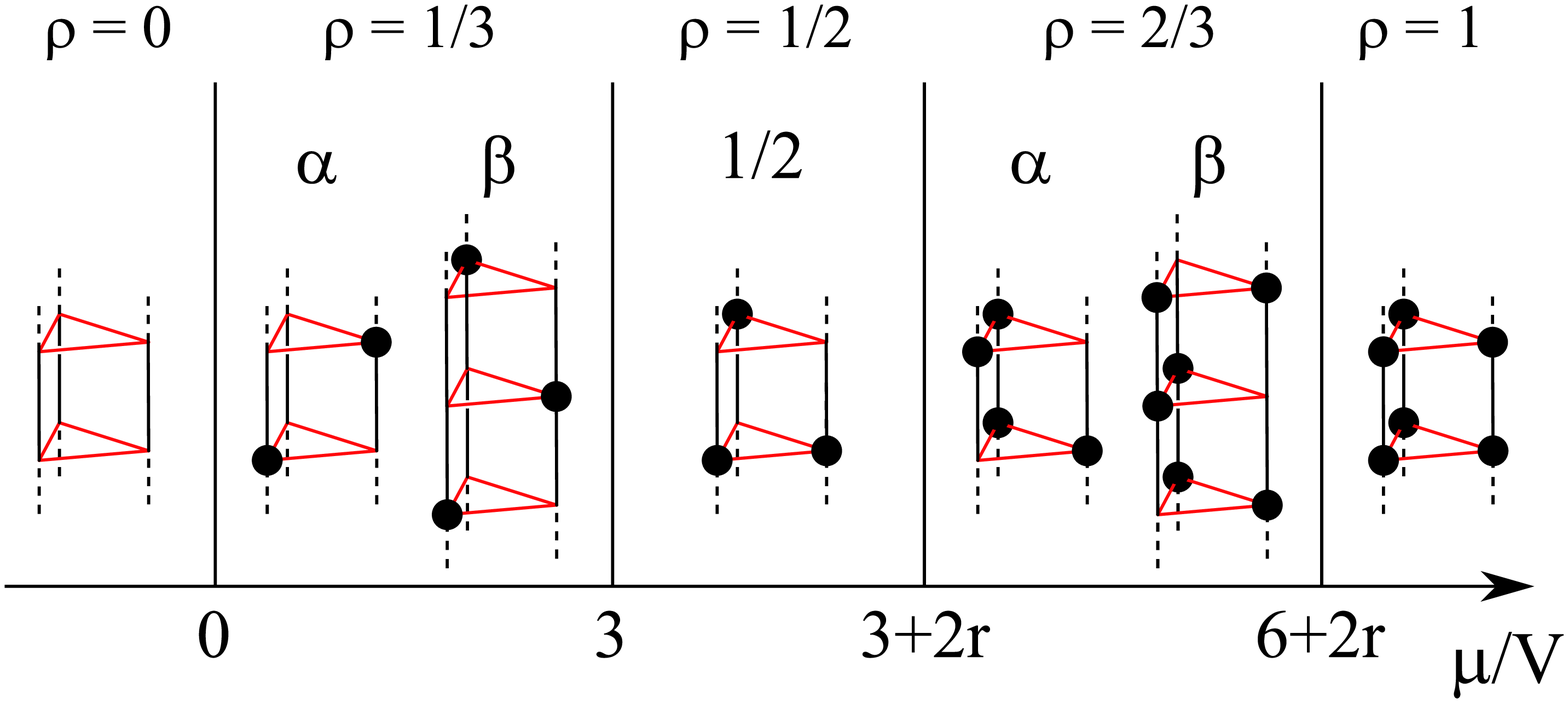}
  \caption{
  The density profiles of solid states realized in strong coupling limit.
  }
  \label{fig:layered-solid}
\end{figure}
In fact, in the simple (classical) limit 
with $t=t'=0$,
a phase diagram should be obtained, as shown in Fig. \ref{fig:layered-solid}.
When $\mu/V <0$ ($6+2r < \mu/V$), an empty (fully-occupied) state appears.
When $0<\mu/V <3$ ($3+2r<\mu/V<6+2r$), the solid state appears with 
$n=1/3$ ($n=2/3$).
If one focuses on a certain layer, the solid state has three-fold degeneracy.
This means that the layered system has a large residual entropy in this limit. 
Here, we specify the solid structure with 
a unit cell composed of six and nine sites 
as $\alpha$- and $\beta$-type solids, respectively 
(see Fig. \ref{fig:layered-solid}).
Note that we did not find any particle distributions
with larger periods when the hoppings $t, t'$ are introduced.
The $\alpha$- and $\beta$-type particle distributions are characterized 
by the following quantities as
\begin{eqnarray}
  \rho_2 &=& \Big|\rho(\frac{4\pi}{3},0,\pi)\Big|,
  \label{structure-factor-S2}\\
  \rho_3 &=& \Big|\rho(\frac{4\pi}{3},0, +\frac{4\pi}{3})\Big| + 
  \Big|\rho(\frac{4\pi}{3},0, -\frac{4\pi}{3})\Big|,
\end{eqnarray}
where
\begin{eqnarray}
  \rho({\bf q})&=&\frac{1}{N}\sum_i \left\langle n_i\right\rangle 
e^{-{\bf q} \cdot {\bf r}_i}.
\end{eqnarray}
On the other hand, at half filling $(n=1/2)$, 
a different particle distribution appears. 
Fig. \ref{fig:layered-solid} shows that the particle density in each layer 
alternate in the $z$-direction.
The state is characterized by the following quantity, 
in addition to $\rho_2$, as
\begin{eqnarray}
  \Delta n = \Big|\rho(0,0,\pi)\Big|.
  \label{structure-factor-delta}
\end{eqnarray}
Here, we specify this particle distribution as the ``$1/2$''-type one, 
for simplicity.
In the weak coupling limit, the superfluid (SF) state is naively expected. 
The order parameter is defined as
\begin{eqnarray}
\Psi &=& \frac{1}{N}\sum_i \langle a_i \rangle.
\end{eqnarray}

When the system has both $\Psi$ and some of the other parameters defined above,
there appears the coexistence between the superfluid and particle density wave,
and we can say that the supersolid state is realized. 
For simplicity, we specify the solid (supersolid) states 
as the S-$\alpha$, S-$\beta$, and S-$1/2$ 
(SS-$\alpha$, SS-$\beta$, and SS-$1/2$), 
corresponding to the particle distributions.
The relation between the ordered states and quantities 
are summarized in Table \ref{tab:solid-order-parameter}.
\begin{table}[htb]
  \centering
  \begin{tabular}{c|ccc|ccc|c}
& S-$\alpha$ & S-$\beta$ & S-$1/2$ & SS-$\alpha$ & SS-$\beta$ & SS-$1/2$ & SF
\\ \hline 
$\rho_2$ & $\neq 0$ & 0 & $\neq 0$ & $\neq 0$ & 0 & $\neq 0$ & 0\\
$\rho_3$ & 0 & $\neq 0$ & 0 & 0 & $\neq 0$ & 0 & 0\\
$\Delta n$ & 0 & 0 & $\neq 0$ & 0 & 0 & $\neq 0$ & 0 \\
$\Psi$ & 0 & 0 & 0 &  $\neq 0$ & $\neq 0$ & $\neq 0$ & $\neq 0$
  \end{tabular}
  \caption{The order parameters of the possible ground states}
  \label{tab:solid-order-parameter}
\end{table}

To discuss ground-state properties in the hard-core Bose-Hubbard
model, we make use of the CMF method~\cite{Oguchi}.
In the method, the original lattice model is mapped 
to an effective cluster model, 
where particle correlations in the cluster
can be taken into account properly. 
The expectation values of the intercluster Hamiltonian are 
obtained via a self-consistency condition imposed on 
the effective cluster problem.
This method has an advantage in discussing quantum phase transitions correctly
since not only stable states but also metastable states can be treated.
Therefore, the CMF method has successfully been applied to 
the quantum spin systems~\cite{Oguchi,Yamamoto2009} and 
bosonic systems~\cite{Hassan,Yamamoto,Suzuki0}.

In the study, we consider the three-leg tube model with the length $L$ as
the effective cluster model, as shown in Fig. \ref{fig:layered-model} (b).
The effective Hamiltonian is explicitly given as,
\begin{eqnarray}
  H_\mathrm{CMF} &=& \sum_C H_\mathrm{CMF}^C, \\
  H_\mathrm{CMF}^C &=&
  - \sum_{\langle i,j \rangle_C} t_{ij}(a_i^\dagger a_j + h.c.)
  + \sum_{\langle i,j \rangle_C} V_{ij}n_i n_j
  \nonumber \\
  &-& \sum_{\langle i,j \rangle'_C} t_{ij}\left( \langle a_j\rangle 
  a_i^\dagger + h.c.\right)
    +\sum_{\langle i,j \rangle'_C} V_{ij}\langle n_j\rangle n_i \nonumber\\
      &-&\mu \sum_{i \in C} n_i +E_0\\
  E_0 &=& \frac{1}{2}\sum_{\langle i,j \rangle'_C} \Big[t_{ij}\Big(\langle a_i\rangle^*\langle a_j\rangle+c.c.\Big)-V_{ij}\langle n_i\rangle\langle n_j\rangle \Big]\nonumber\\
  \label{cmf-hamiltonian}
\end{eqnarray}
where the symbol $\langle i,j\rangle_C$ [$\langle i,k \rangle_C'$] 
denote the summation over nearest neighbor sites in the cluster 
(between clusters).
Note that this three-leg tube is too thin to quantitatively discuss 
quantum phase transitions in the system~\cite{Yamamoto}.
However, 
it is known that an effective three-site cluster in a single layer reproduce
the reasonable phase diagram. 
An important point in this study is that the cluster with a large value of $L$ 
is necessary to discuss the stability of the solid and 
supersolid states with long-period structures.
To this end, we here make use of the DMRG technique 
as an effective cluster solver~\cite{Suzuki0}.
It is known that this method is powerful 
for the one-dimensional systems~\cite{White,DMRG}.
Furthermore, by combining the DMRG method with a mean-field theory,
the phase transitions in higher dimensions have been discussed 
for the Heisenberg models~\cite{Kawaguchi,Suga} and fermionic
Hubbard models~\cite{Maruyama,Garcia}.
Some details of the CMF+DMRG method are explained in Ref.~\cite{Suzuki0}.
In the following, we take the intralayer interaction ($V$) 
as unit of energy and fix the ratio of interactions as $r=1/2$.

\section{Results}\label{4}
In the section, we discuss ground state properties in 
the hard-core Bose-Hubbard model on the layered triangular lattice.
We then clarify the role of the interlayer coupling in stabilizing 
the supersolid state in the system.

\subsection{Low density case}
\begin{figure}[htb]
\includegraphics[width=8cm]{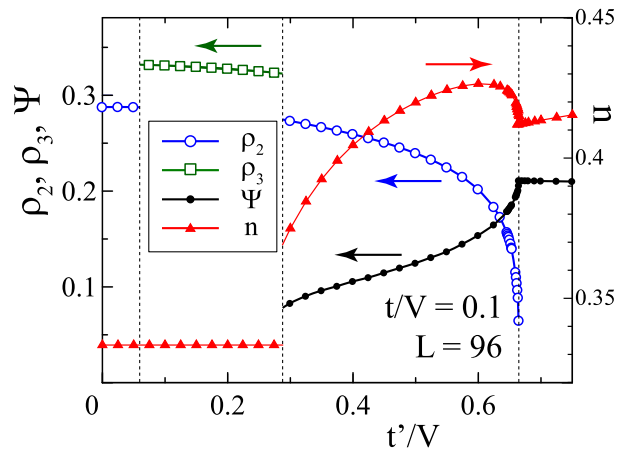}
\caption{
(Color online) The order parameters $\rho_2$ (open circles), 
$\rho_3$ (open squares),
the superfluid order parameter $\Psi$ (solid circles), and
the particle density $n$ (solid triangles)
as a function of $t'/V$
in the system with $t/V=0.1$ and $V'/V=0.5$.
}
\label{fig:low1}
\end{figure}

First, we consider the hard-core bosonic system with a low particle density, 
fixing the chemical potential as $\mu/V=0.68(3+r)$.
Combining the CMF method with the DMRG, 
we obtain the order parameters for the system with a fixed ratio
$t/V=0.1$, as shown in Fig. \ref{fig:low1}.
In this case, 
the particle density in the layer
is always uniform in the $z$-direction ($\Delta n=0$).
When $t'/V=0$, the particle density is one-third ($n=1/3$), 
the order parameter $\rho_2$ is finite and 
the other order parameters are zero. 
This means that the S-$\alpha$ state is realized
in the limit.
The introduction of $t'$ little decreases $\rho_2$ and 
the S-$\alpha$ state still remains.
At $t'/V=(t'/V)_{c1}$,
the particle density is not changed, but $\rho_2$ suddenly vanishes 
and $\rho_3$ appears instead,
where $(t'/V)_{c1}\sim 0.06$.
This implies that the particle distributions are drastically changed, and
the S-$\beta$ state is realized.
Further increase in the hopping 
induces another first-order phase transition at $t'/V=(t'/V)_{c2}$,
where $(t'/V)_{c2}\sim 0.29$.
In the case, both $\rho_2$ and $\Psi$ are induced instead of $\rho_3$
and the particle number becomes away from 
the commensurate filling $(n \neq 1/3)$.
Therefore, we can say that the coexisting state with the 
$\alpha$-type particle distribution (SS-$\alpha$) is realized.
Increasing the hopping $t'$, superfluid correlations are enhanced
and the $\alpha$-type particle distribution becomes obscure, 
as shown in Fig. \ref{fig:low1}.
At $t'/V=(t'/V)_{c3}$, the first-order phase transition 
occurs with small jumps
in the order parameters and the particle density,
where $(t'/V)_{c3}\sim 0.66$.
In the large $t'$ region, no spatial distribution appears
in the particle density
and the genuine superfluid state is realized.

\begin{figure}[htb]
\includegraphics[width=7cm]{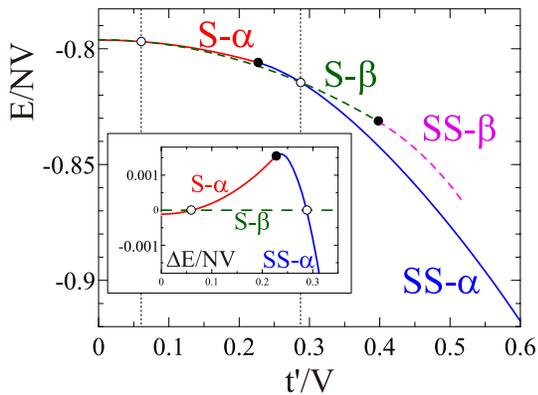}
\caption{
(Color online) The energies 
for S-$\alpha$, S-$\beta$, SS-$\alpha$ and SS-$\beta$ states
in the system with $t/V=0.1$ and $r=0.5$. 
Open circles represent the first-order transition points and 
solid ones represent the second-order transition points in metastable states.
Inset shows the energy difference $\Delta E=E_i-E_{S-\beta}$, where
$i = $S$-\alpha$ and SS-$\alpha$.
}
\label{fig:low2}
\end{figure}

To discuss how the $\alpha$-type state competes with the $\beta$-type one,
we also show the ground state energy in Fig.~\ref{fig:low2}.
The energy for each state is smoothly varied when the interlayer hopping
is varied. 
We find that these curves intersect each other at two points
$(t'/V)_{c1}$ and $(t'/V)_{c2}$,
where the first-order quantum phase transitions occur.
If we focus on the restricted space specified by 
the $\alpha$-type ($\beta$-type) particle distribution,
we find the critical point between the solid and supersolid states 
at $(t'/V)_c\sim 0.23\; (0.40)$,
which is shown as a solid circle in Fig. \ref{fig:low2}.
However, the corresponding energy is higher than the ground-state one,
which means that the quantum phase transition occurs in the metastable state.
Therefore, we could not find critical behavior in the ground state 
when $t/V=0.1$.

\begin{figure}[htb]
\includegraphics[width=7cm]{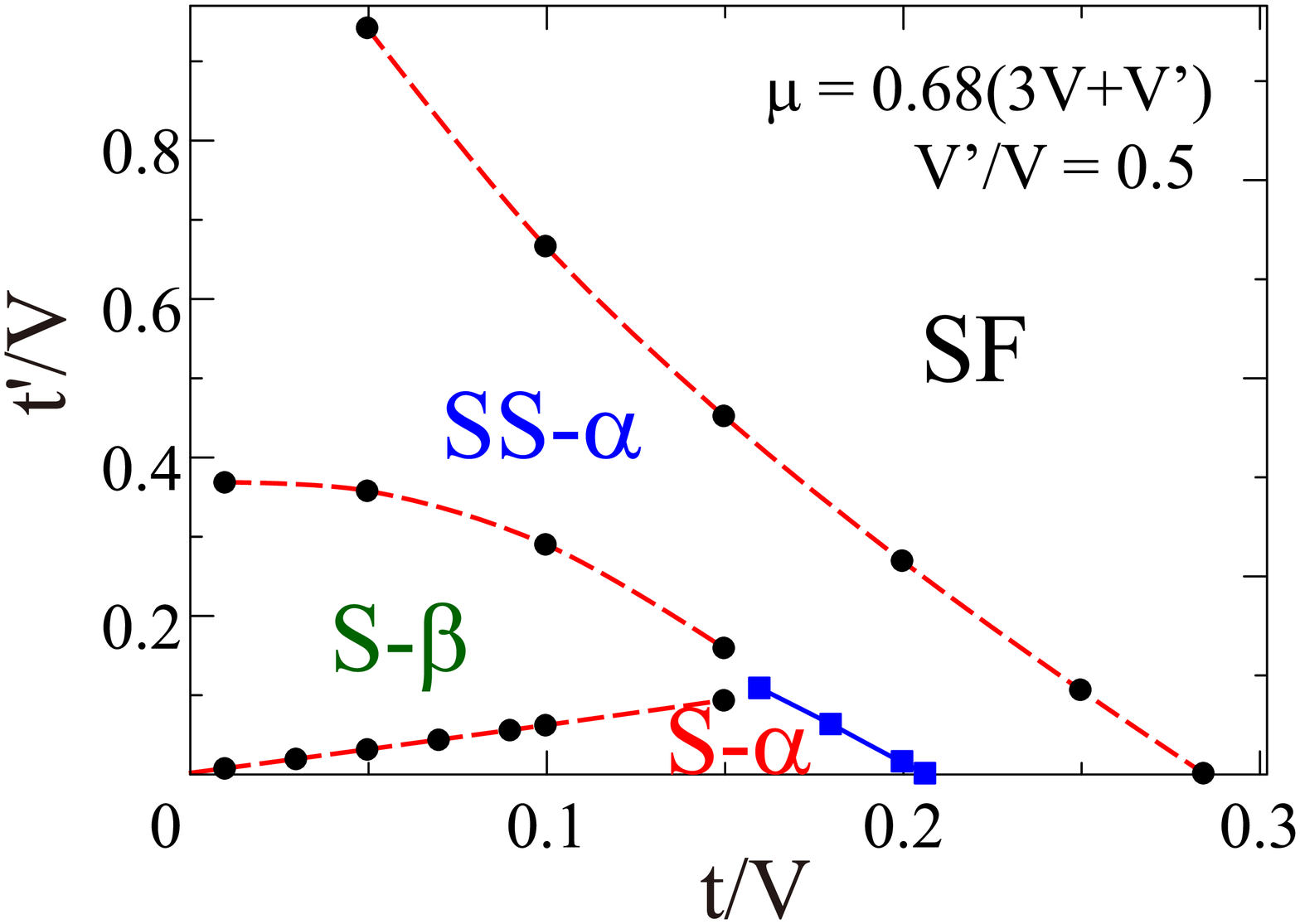}
\caption{
(Color online) The zero temperature phase diagram for the system with
$r=0.5$ and $\mu/V=0.68(3+r)$.
The solid (dashed) lines represent the phase boundary, 
where the first-order (second-order) transition occurs. 
}
\label{fig:lower-filling}
\end{figure}

By performing similar calculations for different values of $t/V$,  
we obtain the zero-temperature phase diagram
as shown in Fig. \ref{fig:lower-filling}.
In the weak coupling region, the superfluid state appears with a finite $\Psi$.
Increasing the interactions ($V, V'$), translational symmetry is broken 
in addition to $U(1)$ symmetry and the $\alpha$-type supersolid state
is realized.
Further increase of the interactions yields 
two solid phases (S-$\alpha$ and S-$\beta$) with $n=1/3$.
We find that the phase transition between the solid and supersolid states 
with the $\alpha$-type particle distribution is 
of second-order and the others are of first-order.

There are two remarkable points in the phase diagram.
One of them is that the S-$\beta$ state is stable against 
the S-$\alpha$ and SS-$\alpha$ states.
Since the S-$\alpha$ and S-$\beta$ states are degenerate in the classical 
limit ($t=t'=0$), 
the competition should be understood by the strong coupling expansion.
The energies for the S-$\alpha$ and S-$\beta$ states are expressed 
by the second-order perturbation theory as
\begin{eqnarray}
  \frac{E_{ S-\alpha}}{NV} &\sim& -\frac{1}{3}\left[  \frac {2}{3+r}\Big(\frac{t'}{V}\Big)^2 + \frac{3}{2}\frac{2+r}{1+r}\Big(\frac{t}{V}\Big)^2 \right],\\
  \frac{E_{ S-\beta}}{NV} &\sim& -\frac{1}{3}\left[  \frac {2}{3}\Big(\frac{t'}{V}\Big)^2 + \frac{6}{2+r}\Big(\frac{t}{V}\Big)^2 \right].
  \label{perturbation-solid-energy}
\end{eqnarray}
This means that the $\alpha$-type solid state is 
mainly stabilized by the intralayer hopping
while the $\beta$-type solid state is by the interlayer hopping.  
The first-order phase transition between two solid states 
occurs at a certain point
\begin{eqnarray}
  (t'/t)_c = \frac 32 \sqrt{ \frac{3r+r^2}{2+3r+r^2} }.
  \label{perturbation-teniten}
\end{eqnarray}
This is consistent with the phase boundary in the phase diagram
although a quantitative difference appears due to a thin tube structure
of the effective cluster. 
For the above reason, 
the S-$\beta$ state is realized in the strong coupling region
with $t'/t>(t'/t)_c$. 
  
\begin{figure}[htb]
\centering
\includegraphics[width=7cm]{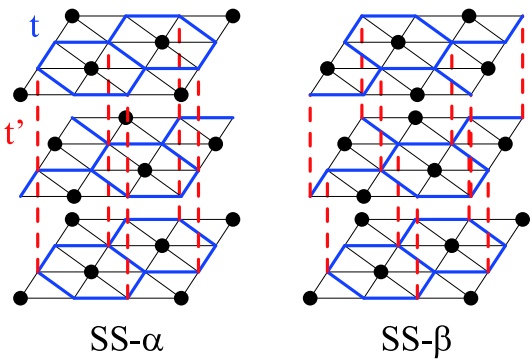}
\caption{
Schematic description of the effective model for the supersolid states 
with $\alpha$ and $\beta$ structures.
Solid circles represent localized particles.
Bold solid and dashed lines represent the hoppings $t$ and $t'$ for the 
defect particles.
}
\label{fig:layered-honeycomb}
\end{figure}

The other point is that 
the SS-$\alpha$ state is always stable 
against the SS-$\beta$ state, in contrast to the competition between 
the solid states discussed above.
This may be explained by considering the nature of the supersolid states.
The supersolid state is the coexistence between 
the solid state with a certain particle distribution and the superfluid state.
Therefore, the stability of the supersolid state with $n > 1/3$ 
is roughly discussed in terms of 
the effective model, which is composed of localized particles and
itinerant ``defect'' particles~\cite{Zhang}.
In this case, the defect particles can hop on layered honeycomb lattices,
as shown in Fig. \ref{fig:layered-honeycomb}.
If the particle gains the kinetic energy against the intersite repulsion, 
particles are spontaneously doped.
This induces the coexistence between the Bose-Einstein condensation 
for defect particles and the solid state for localized particles, that is,
the supersolid state is realized.
By using the first-order perturbation, we obtain the single particle energy 
for each state as
\begin{eqnarray}
  \epsilon_{{SS-\alpha}} &=& -\tilde\mu - t' - \sqrt{(3t)^2 + (t'+V')^2}, \\
  \epsilon_{{SS-\beta}} &=& -\tilde\mu - t' - 3t,
  \label{perturbation-ss-energy}
\end{eqnarray}
where $\tilde \mu = \mu - 3V - V'$.
The relation $ \epsilon_{{SS-\alpha}} < \epsilon_{{SS-\beta}}$ 
is always satisfied, implying that 
the SS-$\alpha$ state is more favorable due to a one-dimensional network 
along the $z$-direction.
This result is consistent with the fact 
that SS-$\beta$ state never appears in the ground-state phase diagram 
(see Fig. \ref{fig:lower-filling}).
We also find that the phase boundary 
between the SS-$\alpha$ and S-$\alpha$ states
roughly obtained from $\epsilon_{SS-\alpha}=0$ is consistent with 
that in the phase diagram (not shown).

In the section, we have clarified that the SS-$\alpha$ state 
is indeed realized at the low particle density.
As discussed above, the nature of the SS-$\alpha$ state is essentially 
the same as that of the supersolid state in the single layered model.
Therefore, we can say that the supersolid state realized 
in the single layer is 
stable against the interlayer coupling.

\subsection{Half filling}
\begin{figure}[htb]
\includegraphics[width=7cm]{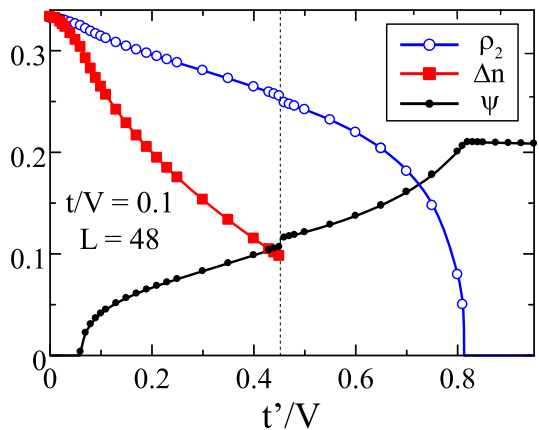}
\caption{
(Color online) 
The order parameters $\rho_2$ (open circles), 
$\Delta n$ (solid squares), and
the superfluid order parameter $\Psi$ (solid circles) as a function of $t'/V$
in the system with $t/V=0.1, r=0.5$ and $\mu/V=3+r$.
}
\label{fig:half1}
\end{figure}

Here, we consider the hard-core bosonic system at the symmetric condition
with $\mu/V=3+r$ to clarify how the supersolid state
 is stabilized.
Fig. \ref{fig:half1} shows the order parameters for the system 
with a fixed ratio $t/V=0.1$.
In this case, the system is always half-filled ($n=1/2$) and $\rho_3$ is zero.
Therefore, the $\beta$-type particle distribution
is not realized at half filling.
When $t'/V<(t'/V)_{c1}$, $\rho_2$ and $\Delta n$ are finite and $\Psi=0$,
where $(t'/V)_{c1}\sim 0.06$. 
This implies that the S-$1/2$ state is realized.
Increasing the hopping $t'$, 
the second-order phase transition occurs to the SS-$1/2$ state
at $t'/V=(t'/V)_{c1}$, where $\Psi$ is induced.
A further increase in the hopping $t'$ increases $\Psi$ and 
decreases $\rho_2$ and $\Delta n$.
At the transition point, 
$\Delta n$ suddenly vanishes and the phase transition occurs 
to the SS-$\alpha$ state, where the small jumps appear
in the curves of $\rho_2$ and $\Psi$.
The first-order transition point is determined as $(t'/V)_{c2}=0.45$,
by deducing the crossing point of the energy curves for competing states
(not shown).
The increase in the hopping $t'$ decreases $\rho_2$ and increases $\Psi$.
Finally, $\rho_2$ vanishes and the second-order phase transition occurs 
to the superfluid state at a critical point $(t'/V)_{c3}$,
where $(t'/V)_{c3}\sim 0.81$.

\begin{figure}[htb]
\includegraphics[width=7cm]{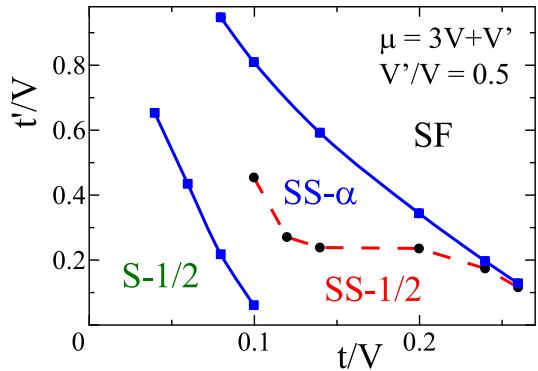}
\caption{(Color online)
The zero temperature phase diagram for the half filled system with $r=0.5$. 
The solid (dashed) lines represent the phase boundary, where 
the second- (first-)order phase transition occurs. 
}
\label{fig:half3}
\end{figure}
We perform the CMF calculations for several values of $t/V$ 
and end up with the phase diagram at half filling,
as shown in Fig. \ref{fig:half3}.
In the strong coupling region, the S-$1/2$ state is realized, where
alternating behavior appears in the local particle density.
The increase in the hoppings $t$ and $t'$ enhances superfluid correlations and 
the quantum phase transition occurs.
The SS-$1/2$, SS-$\alpha$, and superfluid states are realized, 
depending on the parameters.
When $t/V$ is small and $t'/V$ is large, 
the obtained order parameters strongly depend on 
the tube length $L$ in the CMF+DMRG method, and 
thereby we could not determine the phase boundaries.

What is the most important is that the supersolid states are widely realized 
between the superfluid and solid states.
This is contrast to the results discussed 
in the single layered model
where the supersolid state is not stable at half filling.
To clarify how the SS-$1/2$ and SS-$\alpha$ states 
are stabilized at half filling,
\begin{figure}[htb]
  \centering
  \includegraphics[width=8cm]{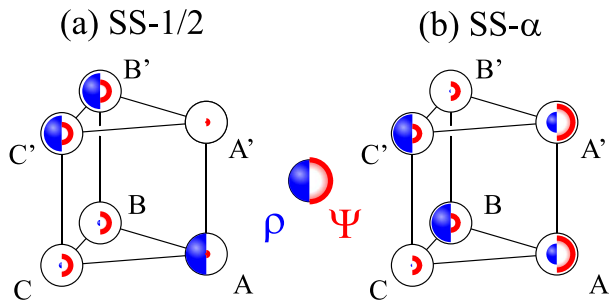}
  \caption{(Color online)
  The distributions of the order parameters at the sublattices 
  A, B, C, A', B', and C' in the half-filled system 
  with $t/V = 0.1$ when $t'/V = 0.3$ (a) and 0.5 (b). 
  Radius of left and right semicircles at each sublattice represent 
  a local particle density $\langle n_{im}\rangle$ and 
  a local superfluid order parameter $\Psi$, respectively. 
  }
  \label{fig:sshf-density}
\end{figure}
we show local particle density and superfluid order parameter 
in Fig. \ref{fig:sshf-density}.
In the SS-$1/2$ state $(t/V=0.1$ and $t'/V=0.3)$,
the superfluid order parameter is finite at each site. 
We also find that most of bosons are located at sublattice A in a certain layer
and at sublattices B' and C' in a different one.
This feature recalls us the supersolid state in the single layered model 
away from half filling.
In fact, the SS-$1/2$ state may be regarded as the supersolid state with 
alternating layered structure, where the supersolid state with $n<1/2(>1/2)$ is
realized in the even (odd) layers.
Therefore, we can say that 
this SS-$1/2$ state in the layered model is stabilized 
by the self-doping effect
through the interlayer hopping.
In the other word, the translational symmetry 
for the particle density in each layer is spontaneously broken,
which stabilizes the SS-$1/2$ state at half filling.

On the other hand, a different mechanism stabilizes the SS-$\alpha$ state.
When $t/V=0.1$ and $t'/V=0.5$, the SS-$\alpha$ state is realized with 
three distinct sublattices in each layer, 
in contrast to the SS-$1/2$ state, as shown in Fig. \ref{fig:sshf-density}.
At the sublattices A and A', the local particle density is 
close to half filling and the other sublattices are almost 
empty or fully-occupied.
This enhances superfluid correlations along the chains 
through the sublattices A and A'. 
In fact, the order parameter at each sublattice is maximum 
and its proximity effect stabilizes the superfluid state in the whole system.
On the other hand, the intersite interactions $(V, V')$ stabilize 
the alternating structure in the local particle density,
where B and C' (B' and C) sublattices are almost fully-occupied (empty),
as shown in Fig. \ref{fig:sshf-density} (b).
Therefore, we can say that the SS-$\alpha$ state at half filling
is mainly stabilized by the one-dimensional network in $z$-direction.

Before closing the section, we would like to comment on 
experiments on the optical lattice. 
It is known that the supersolid state in a single layered model
has three-fold degeneracy. 
Therefore, the interlayer coupling should play a crucial role
in observing the supersolid state in the layered triangular lattice.
In the study, we have clarified that the supersolid states
are widely stable at not only the incommensurate fillings
but also half filling.
Therefore, we can say that
ultracold bosons on the layered triangular lattice 
is one of the most appropriate systems to observe the supersolid states.

\section{Summary}

We have studied ground-state properties in the hard-core Bose-Hubbard model 
on a layered triangular lattice. 
Combining cluster mean-field theory with the density matrix 
renormalization group method, 
we have discussed how stable the supersolid states realized 
in a single layer triangular lattice are against the interlayer coupling.
We have obtained the rich phase diagram of the system.
It has been found that two types of supersolid states are realized 
even at the commensurate filling.

\section{Acknowledgments}
The authors would like to thank I. Danshita and D. Yamamoto 
for valuable discussions.
This work was partly supported by Japan Society for the Promotion of Science 
Grants-in-Aid for Scientific Research Grant Number 25800193
and the Global COE Program ``Nanoscience and Quantum Physics" from 
the Ministry of Education, Culture, Sports, Science and Technology (MEXT) 
of Japan. 


\end{document}